\documentclass{article}

\usepackage{arxiv}

\usepackage[utf8]{inputenc} 
\usepackage[T1]{fontenc}    
\usepackage{lmodern}        
\usepackage{hyperref}       
\usepackage{url}            
\usepackage{booktabs}       
\usepackage{amsfonts}       
\usepackage{nicefrac}       
\usepackage{microtype}      
\usepackage{graphicx}

\title{Automatic Scoring of Cognition Drawings}

\author{
    Arne Bethmann
   \\
    SHARE Germany\\
MEA-SHARE and \\
  SHARE Berlin Institute \\
  \texttt{\href{mailto:abethmann@share-berlin.eu}{\nolinkurl{abethmann@share-berlin.eu}}} \\
   \And
    Marina Aoki
   \\
    SHARE Germany\\
MEA-SHARE and \\
  SHARE Berlin Institute \\
  \texttt{\href{mailto:maoki@share-berlin.eu}{\nolinkurl{maoki@share-berlin.eu}}} \\
   \And
    Charlotte Hunsicker
   \\
    SHARE Germany\\
MEA-SHARE and \\
  SHARE Berlin Institute \\
  \texttt{\href{mailto:chunsicker@share-berlin.eu}{\nolinkurl{chunsicker@share-berlin.eu}}} \\
   \And
    Claudia Weileder
   \\
    SHARE Germany\\
MEA-SHARE and \\
  SHARE Berlin Institute \\
  \texttt{\href{mailto:cweileder@share-berlin.eu}{\nolinkurl{cweileder@share-berlin.eu}}} \\
  }

\providecommand{\tightlist}{%
  \setlength{\itemsep}{0pt}\setlength{\parskip}{0pt}}

\newlength{\cslhangindent}
\newlength{\csllabelwidth}
\newlength{\cslentryspacingunit} 
  {}%
  {\par}
\newenvironment{CSLReferences}[2] 
 {
  \setlength{\parindent}{0pt}
  \ifodd #1
  \let\oldpar\par
  \def\par{\hangindent=\cslhangindent\oldpar}
  \fi
  \setlength{\parskip}{#2\cslentryspacingunit}
 }%
 {}
\usepackage{calc}

\usepackage{setspace,booktabs,url,soul,color,hyperref,float}
\sethlcolor{yellow}
\hypersetup{ colorlinks=false, pdfborder={0 0 0}, }
\begin{document}
\maketitle

\begin{abstract}
Figure drawing is often used as part of dementia screening protocols.
The Survey of Health Aging and Retirement in Europe (SHARE) has adopted
three drawing tests from Addenbrooke's Cognitive Examination III as part
of its questionnaire module on cognition. While the drawings are usually
scored by trained clinicians, SHARE uses the face-to-face interviewers
who conduct the interviews to score the drawings during fieldwork. This
may pose a risk to data quality, as interviewers may be less consistent
in their scoring and more likely to make errors due to their lack of
clinical training. This paper therefore reports a first proof of concept
and evaluates the feasibility of automating scoring using deep learning.
We train several different convolutional neural network (CNN) models
using about 2,000 drawings from the 8th wave of the SHARE panel in
Germany and the corresponding interviewer scores, as well as
self-developed `gold standard' scores. The results suggest that this
approach is indeed feasible. Compared to training on interviewer scores,
models trained on the gold standard data improve prediction accuracy by
about 10 percentage points. The best performing model, ConvNeXt Base,
achieves an accuracy of about 85\%, which is 5 percentage points higher
than the accuracy of the interviewers. While this is a promising result,
the models still struggle to score partially correct drawings, which are
also problematic for interviewers. This suggests that more and better
training data is needed to achieve production-level prediction accuracy.
We therefore discuss possible next steps to improve the quality and
quantity of training examples.
\end{abstract}


\hypertarget{introduction}{%
\section{Introduction}\label{introduction}}

Dementia and related disorders are among the leading causes of death,
disability and dependency among the global elderly population (World
Health Organization, 2017). In addition to the severe cognitive, mental
and physical impact on individuals with dementia, the disease also
affects their social networks and, with increasing prevalence, society
as a whole. The estimated number of cases is expected to increase from
around 55 million in 2021 to 78 million in 2030 and 139 million in 2050
(World Health Organization, 2021), largely due to the ageing of the
world's population. There are currently no known cures for the disease,
making the identification and management of risk factors one of the few
promising approaches to address this emerging crisis.

Cognitive tests such as the Montreal Cognitive Assessment (MoCA, see
Nasreddine et al., 2005) or the Mini-Mental State Examination (MMSE, see
Folstein et al., 1975) can be used to detect early signs of cognitive
decline. The correct drawing of figures such as cubes or clocks is often
used as an indicator in these tests to assess the visuospatial domain of
cognitive functioning. They are usually administered in a clinical
setting and used in conjunction with additional assessments to diagnose
cognitive impairment, such as dementia-type disorders. The drawings are
usually scored by trained clinicians.

More recently, cognitive drawing tests have been used in several
large-scale survey studies, such as SHARE, the Survey of Health, Ageing
and Retirement in Europe\footnote{\url{https://share-eric.eu/}} in its
main survey, and the HCAP network studies (Langa et al.,
2019).\footnote{\url{https://hcap.isr.umich.edu/}} While these tests are
adapted from clinical cognitive assessments, in this setting they are
used to make population-based estimates, such as the prevalence of
dementia symptoms, rather than individual diagnoses. This will allow
researchers to look for patterns in early cognitive decline and
hopefully identify factors that can slow or delay its progression.

Collecting this data on a large scale in an interviewer-administered
survey context means, in the case of SHARE, tens of thousands of
interviews conducted repeatedly by thousands of interviewers over
several waves. This requires a high degree of standardisation to ensure
consistent measurement, which is particularly challenging given that
survey interviewers are generally not clinically trained to diagnose
cognitive problems.

Since wave 8 (2019/2020), SHARE has included three drawing tests from
Addenbrooke's Cognitive Examination III (see Wagner \& Douhou, 2021) as
part of the cognitive impairment questionnaire module. To date, SHARE
has relied on regular face-to-face interviewers to carry out scoring
during fieldwork. Training procedures and office-based scoring have
shown that this is not a trivial task and that it warrants closer
scrutiny and, ideally, a more standardised approach to mitigate data
quality issues.

In this paper, we explore the use of machine learning, specifically
convolutional neural networks, to automate cognitive drawing scoring,
with the aim of improving data quality and reducing the burden of
scoring on interviewers. Our approach involves training models such as
ConvNeXt (Liu et al., 2022), AlexNet (Krizhevsky et al., 2012), VGG
(Simonyan \& Zisserman, 2014) and ResNet50 (He et al., 2015) on two
types of datasets: one scored by field interviewers and another using a
self-developed `gold standard' procedure. We first assess the ability of
the models to learn from interviewer-provided scores, and then examine
the improvement in accuracy when these models are trained on the more
consistently scored gold standard data. We then compare the accuracy of
manual interviewer scoring with model-based scores trained on either
dataset. Subsequently, we look at differences in model performance due
to different hyperparameter combinations. Finally, we consider the
extent and nature of errors produced by the models and suggest how these
might be mitigated in further research. This comprehensive analysis aims
to demonstrate the potential of deep learning to standardise and improve
the accuracy of cognitive assessment scoring in large-scale survey
studies.

\hypertarget{related-research}{%
\section{Related Research}\label{related-research}}

Interviewer effects have long been recognised as a potential source of
error in survey research and are therefore well studied (see West \&
Blom, 2016). There is no comprehensive research on interviewer effects
in cognitive tests in surveys, but there is reason to believe that
interviewers may also influence the data collection process in these
areas (Malhotra et al., 2015). Despite additional training, interviewers
may face unique challenges when tasked with cognitive tests, as these
assessments are not part of their routine responsibilities.

Initial applications of machine learning to cognitive testing have been
conducted in clinical neuropsychology to aid in the diagnosis of
cognitive impairment (Binaco et al., 2020; Youn et al., 2021). The
datasets used in these studies are not only comparatively small, but
also very specific in their composition, often including a large
proportion of individuals suspected of having some degree of cognitive
impairment.

There have also been efforts to use machine learning to classify
drawings from cognitive assessments collected in representative surveys.
Hu et al. (2022) used data from the National Health and Aging Trends
Study to train a machine learning model to score drawings from the Clock
Drawing Test (CDT). Similarly, Amini et al. (2021) trained a machine
learning model to predict a patient's dementia status based on drawings
from the CDT administered in the Framingham Heart Study.

\hypertarget{data-methodology}{%
\section{Data \& Methodology}\label{data-methodology}}

In this paper we use data from the eighth wave of the Survey of Health,
Ageing and Retirement in Europe (SHARE, Börsch-Supan, 2022).\footnote{The
  interviewer scores and other data from the survey interviews are part
  of the SHARE data release, while the drawings used in this paper are
  not yet published.} SHARE is a biennial panel survey conducted in 28
European countries: all EU Member States except Ireland, plus Israel and
Switzerland. The target population is people aged 50 and over, who are
asked a wide range of questions about their socio-economic situation,
social networks and health. To date, some 600,000 interviews have been
collected from around 150,000 respondents in nine waves of the survey.
The data are available to the scientific community free of
charge.\footnote{\url{https://share-eric.eu/data/data-access}}

The actual drawings for the cognitive tests\footnote{see questions CF830
  to CF839 in the SHARE generic CAPI questionnaire wave 8:
  \url{https://share-eric.eu/fileadmin/user_upload/Questionnaires/Q-Wave_8/paperverstion_en_GB_8_2_5b.pdf}}
are made in a recording booklet, which is usually discarded after
fieldwork. For this study, we collected the recording booklets from the
German SHARE sub-study. Out of a total of 2,878 respondents, 2,374 were
asked to complete the drawing tests in the cognitive module of the
questionnaire because they were over 60 at the time of the interview. We
retrieved and scanned 2,109 booklets.

\hypertarget{cognitive-functioning-in-share}{%
\subsection{Cognitive functioning in
SHARE}\label{cognitive-functioning-in-share}}

\begin{figure}

{\centering \includegraphics[width=0.75\linewidth]{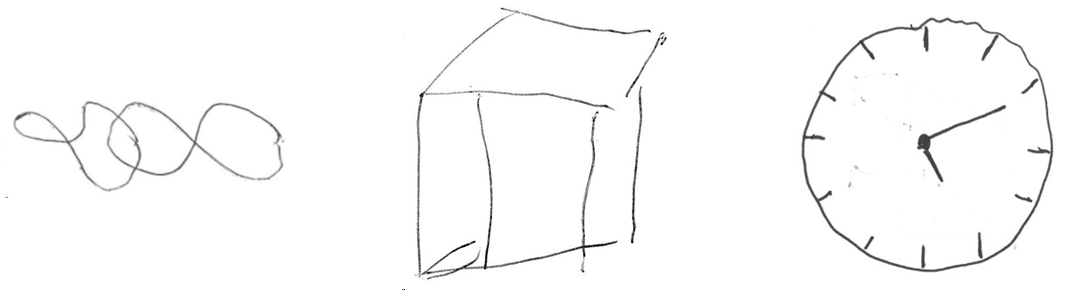} 

}

\caption{Examples of the three drawings: two overlapping infinity loops, a three-dimensional cube, and a clock with its hands set to a specific time.}\label{fig:drawings}
\end{figure}

Since its inception, SHARE has had a strong focus on health topics, both
physical and mental. For example, items on cognitive functioning have
been included since wave 1 (Börsch-Supan \& Jürges, 2005, pp. 182--186)
and have been collected continuously since then. In wave 8, new
cognitive measures were introduced in SHARE, including for the first
time tests from Addenbrooke's Cognitive Examination III (ACE-III, see
Hsieh et al., 2013).\footnote{More recently, the Harmonised Cognitive
  Assessment Protocol (HCAP, see
  \url{https://share-eric.eu/data/data-set-details/share-hcap}) was
  introduced, an additional in-depth study of cognitive functioning
  carried out in some of the SHARE countries.} These aim to assess a
respondent's abilities in the visuospatial domain of cognitive
functioning by asking them to draw three figures: two overlapping
infinity loops, a three-dimensional cube, and a clock with its hands set
to a specific time (see examples in Figure \ref{fig:drawings}). The
loops and the cube had to be copied from a printed picture directly
above the drawing area, while the clock had to be drawn from memory (see
Wagner \& Douhou, 2021, p. 44). These drawings were scored by the
interviewers during the interview and the result was recorded in the
CAPI questionnaire. For the analyses in this paper, we will focus on the
cube drawings. The other drawings will form part of subsequent research.

\hypertarget{interviewer-scoring}{%
\subsection{Interviewer scoring}\label{interviewer-scoring}}

As part of the fieldwork preparations for each SHARE wave, interviewers
receive two days of comprehensive training. During this programme, part
of the time is devoted to instructing interviewers in the administration
of the cognitive module of the questionnaire. With regard to the scoring
of each drawing, the aim was to facilitate the interviewer's
understanding of the specified scoring criteria and to ensure a clear
and consistent scoring process. The training materials included a
detailed description of the scoring rules as well as a set of sample
drawings with scores. The scoring rules for the cube drawing were as
follows

\begin{figure}

{\centering \includegraphics[width=0.5\linewidth]{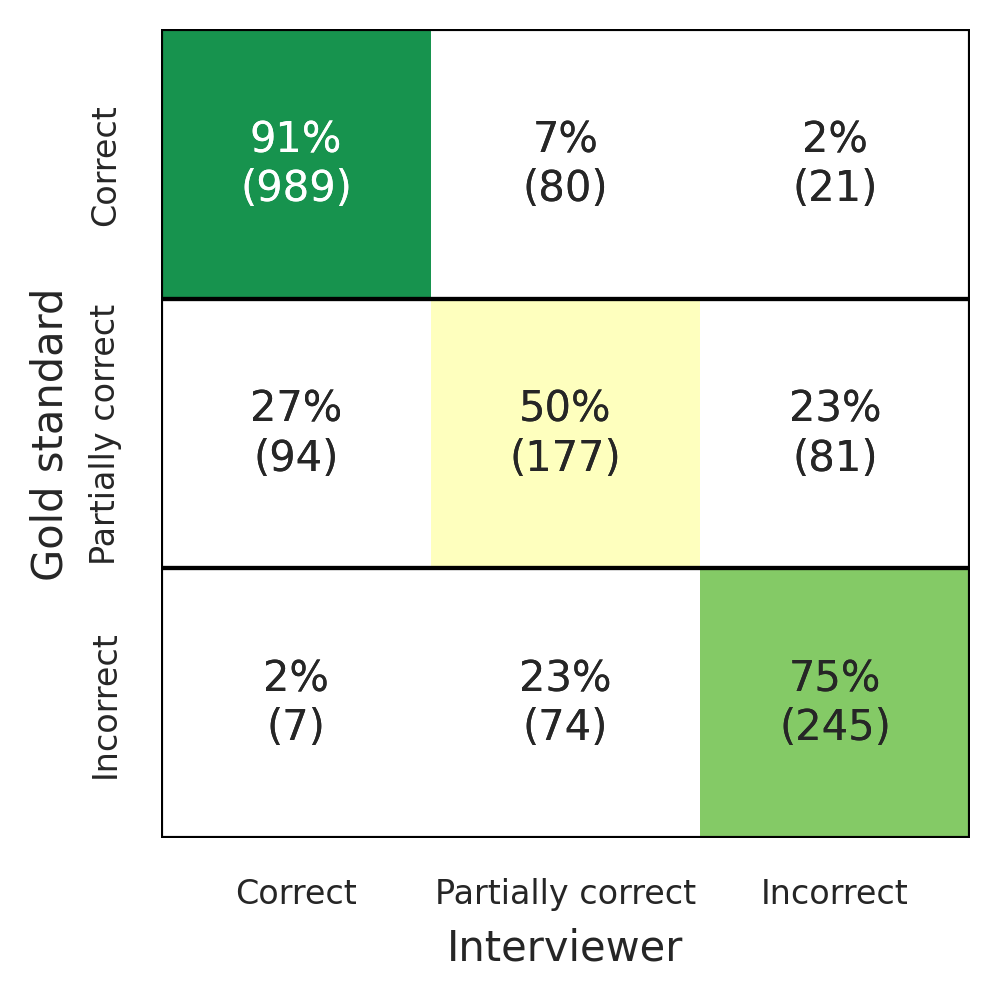} 

}

\caption{Confusion matrix of interviewer scores against gold standard scores. The percentages in the rows show the proportion of interviewer scores in each class, with correct scores highlighted. Absolute numbers are given in parentheses.}\label{fig:gsintc}
\end{figure}

\begin{itemize}
\tightlist
\item
  Fully correct copy: the cube has 12 lines, even if the proportions are
  not perfect.
\item
  Partially correct copy: the cube has fewer than 12 lines, but the
  general shape of the cube is maintained.
\item
  Incorrect copy
\end{itemize}

Respondents were allowed to correct mistakes during the drawing or to
try again if they wished, but interviewers were not allowed to give
feedback or otherwise help respondents with the drawing. Finally, the
drawing of the cube was to be scored by the interviewer immediately
after completion. Of the 2,109 recording booklets collected and
processed, 1,769 contained usable cube drawings with interviewer scores.

\hypertarget{gold-standard-scoring}{%
\subsection{Gold standard scoring}\label{gold-standard-scoring}}

As already known from the literature (e.g. Plank, 2022; Say \&
O'Driscoll, 2022) and confirmed during the interviewer training sessions
and scoring tests in the office, it is clear that scoring the cube
drawings is not as straightforward as it might seem. The scoring rules
are not always clear and there are many borderline cases. In addition,
the interviewers are not clinically trained and may not be able to
identify certain types of error. This is particularly true of the
partially correct category, where the cube shape is maintained but some
lines are missing. For example, it is not always clear whether a line is
missing or simply not visible due to the drawing technique. In addition,
the interviewers are familiar with the respondents and may include
factors other than just looking at the drawing in their judgement and
final score. To reduce this measurement error, we also introduce
in-house scoring, which we will use as a benchmark and refer to as `gold
standard' or `ground truth' scores (depending on the scientific
discipline; for the sake of simplicity, we will use `gold standard' for
the remainder of the paper). In this process, the scorers are given only
the drawings, with no additional information about the respondents, and
are asked to score according to the same set of rules as the
interviewers.

The gold standard scoring was carried out by three scorers, two of whom
had previous experience in scoring drawings from cognitive tests, albeit
in the context of the HCAP study, which has a different set of rules.
The third scorer had no previous experience. All three were trained
in-house using the same materials as the interviewers. Each cube was
scored independently by all three scorers. The scorers were instructed
to score the drawings according to the same rules as the interviewers,
but without access to the previous scores. In cases of disagreement
between the scores, including those of the interviewers, an arbitration
round was added in which each case was discussed and a final scoring
decision was made, by majority vote if necessary. This was done by a
team of two of the scorers and a third person. In total, 1,776 cubes
were scored by the scorers, slightly more than by the interviewers, as
some usable drawings with missing interviewer scores could be recovered.
905 of these had to be decided by the arbitration team. The final gold
standard scores were then used to train the models.

A comparison of the interviewer scores with the gold standard shows an
accuracy rate of 79.8\% (Figure \ref{fig:gsintc}). The class percentages
in the gold standard are: `correct' 61.65\%, `partially correct' 19.91\%
and `incorrect' 18.44\%, so quite unbalanced with about 60/20/20.
Scoring correct cubes was the easiest for the interviewers, leading to
91\% correct scores in this class. Identifying incorrect cubes was more
difficult, but interviewers still achieved an accuracy rate of 75\%. The
partially correct cubes are indeed problematic, and interviewers
achieved only 50\% accuracy, still better than chance in our particular
3-class problem, but not very good either.

\begin{table}
\centering
\caption{Percentage of respondents with a self-reported diagnosis of dementia.}
\label{tab:diag}
\vspace{1em}
\begin{tabular}{lcc}
\toprule
Drawing score & Interviewer & Gold standard \\
\midrule
Correct           & 2.57 \% & 2.52 \% \\
Parially correct  & 1.82 \% & 2.09 \% \\
Incorrect         & 5.50 \% & 5.61 \% \\
\bottomrule
\end{tabular}
\end{table}

Table \ref{tab:diag} relates the score to the respondent's self-reported
dementia diagnosis as a plausibility check. Although the number of
respondents with self-reported dementia is very small, there is some
association with incorrectly drawn cubes as opposed to correct cubes,
which is what we would hope to see with an indicator in a dementia
screening test. Partially correct cubes are actually less likely to be
associated with reported dementia than fully correct cubes. This seems a
bit strange, but given the small sample size it may just be due to low
statistical power. We also see that there does not seem to be a
substantial difference between these patterns of association for the
interviewer scores and the gold standard scores.

\hypertarget{models}{%
\subsection{Models}\label{models}}

We trained several different computer vision deep learning models to
predict the class of cubes drawn by respondents as either `correct',
`partially correct' or `incorrect'. For the actual training datasets,
the drawings were scanned, cropped, resized to 128 by 128 pixels and
converted to grey scale. Where this could not be done automatically,
manual corrections were made. The training/validation dataset split was
75\%/25\%, i.e.~the actual training dataset size was 1,332 out of a
total of 1,776 cases with gold standard labels.

As a general baseline, we used AlexNet (Krizhevsky et al., 2012) and
then tested ResNet50 (He et al., 2015) and VGG19\_BN (Simonyan \&
Zisserman, 2014) as intermediate, commonly used models. Finally, we also
trained more recent ConvNeXt models (Small, Base and Large, see Liu et
al., 2022), which have shown competitive performance in image
classification tasks\footnote{see
  e.g.~\url{https://github.com/huggingface/pytorch-image-models/blob/main/results/README.md}}
and which we expected to perform well for our classification problem.
All models used were Convolutional Neural Networks (CNN), although we
also considered using Vision Transformers (ViT, Dosovitskiy et al.,
2020). However, as ViTs are considered to be most useful on larger
datasets, we decided to stick with CNNs for the time being. We will
revisit this decision in future research.

\begin{figure}
\includegraphics[width=1\linewidth]{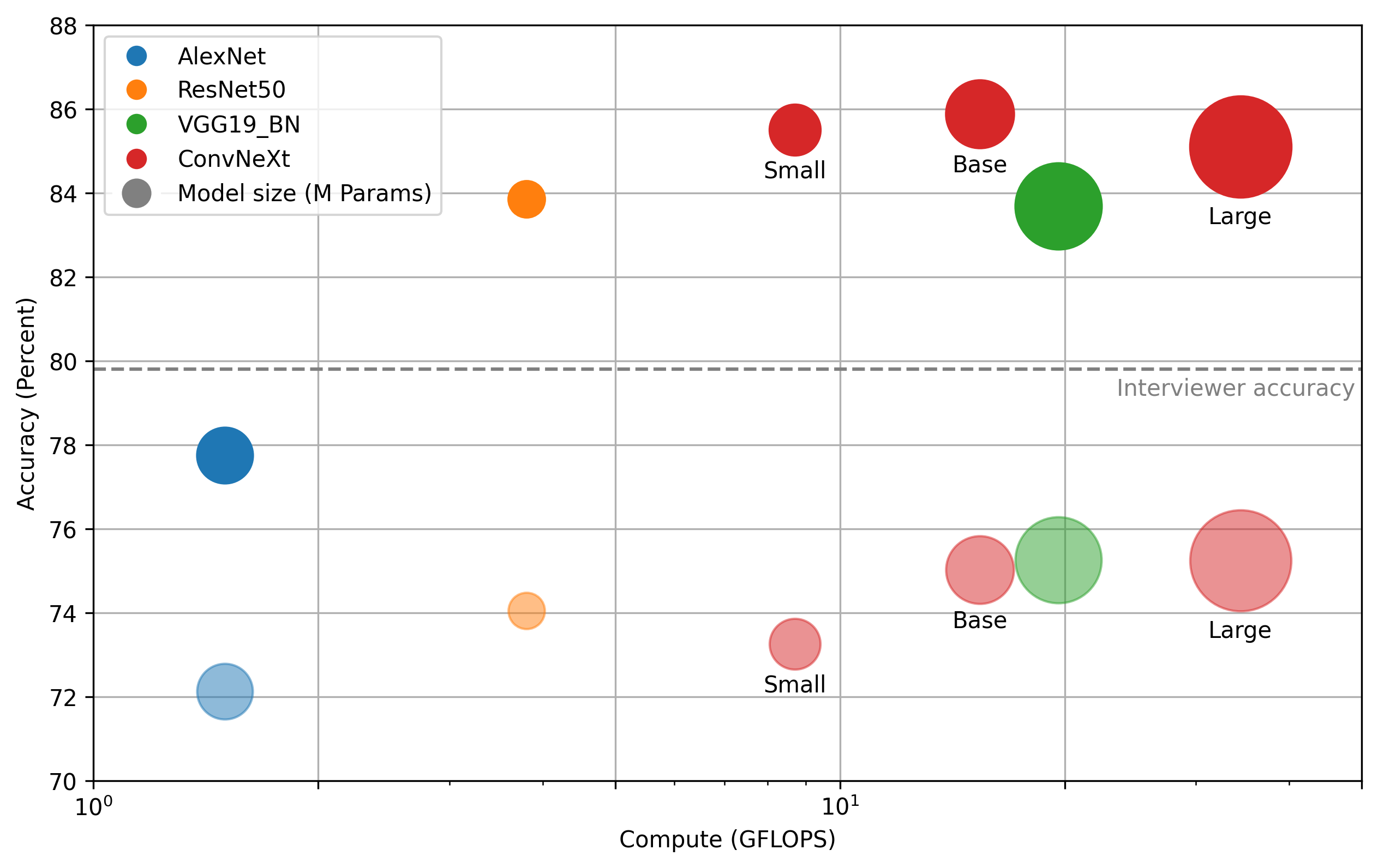} \caption{Performance of different model architectures on interviewer scored labels (semi-transparent) and gold standard scored labels (solid). Mean accuracy values across all experiments.}\label{fig:model-comp}
\end{figure}

\hypertarget{experiments}{%
\subsection{Experiments}\label{experiments}}

For each model, we ran a series of fine-tuning experiments on both
training datasets, the one labelled with the interviewer scores and the
one labelled with the gold standard scores. We used models pre-trained
on the ImageNet data (Deng et al., 2009). The hyperparameters we varied
in the experiments were training duration (50 vs.~100 epochs) and data
augmentation (none, transform only, transform plus resize). Each
experiment was run three times with different seeds, which were the same
for all combinations of hyperparameters. In total, we ran 216
experiments, which became part of our analysis. Model training was
performed using the fastai (Howard \& Gugger, 2020) and timm (Wightman,
2019) libraries. All experiments were run on an NVIDIA RTX A5000 in a
Lenovo P360 workstation with 64 GB of RAM.

\hypertarget{results}{%
\section{Results}\label{results}}

Figure \ref{fig:model-comp}, clearly shows that using our gold standard
procedure to label the training data substantially increases prediction
accuracy across all model architectures (compare solid circles for gold
standard scores vs.~semi-transparent circles for interviewer scores).
This is likely due to more consistent scoring, i.e.~similar looking
cubes are less ambiguously scored as the same class. Whether this is
actually an indicator of higher quality labels needs to be evaluated by
experts such as clinical psychologists. For now, we accept this
assumption and see that the increase in accuracy is in the order of ten
percentage points for all models except AlexNet. It is also the only
model that does not beat the accuracy of human interviewers, which is
close to 80\%.

An interesting observation is that the difference between the models is
more pronounced with higher quality training data. While ConvNeXt
performs similarly to ResNet50 and VGG19\_BN on the lower quality data,
it outperforms all other models when trained on the gold standard data.
The amount of computation required is clearly, but not perfectly,
related to performance. In our experiments, ConvNeXt Base performed
best, even though it is both smaller and less computationally intensive
than VGG19\_BN and ConvNeXt Large. Even the smaller and less
computationally intensive models are competitive. ConvNeXt Small comes
second in terms of accuracy, and even ResNet50 is only beaten by the
much newer ConvNeXt family of models.

\begin{figure}

{\centering \includegraphics[width=1\linewidth]{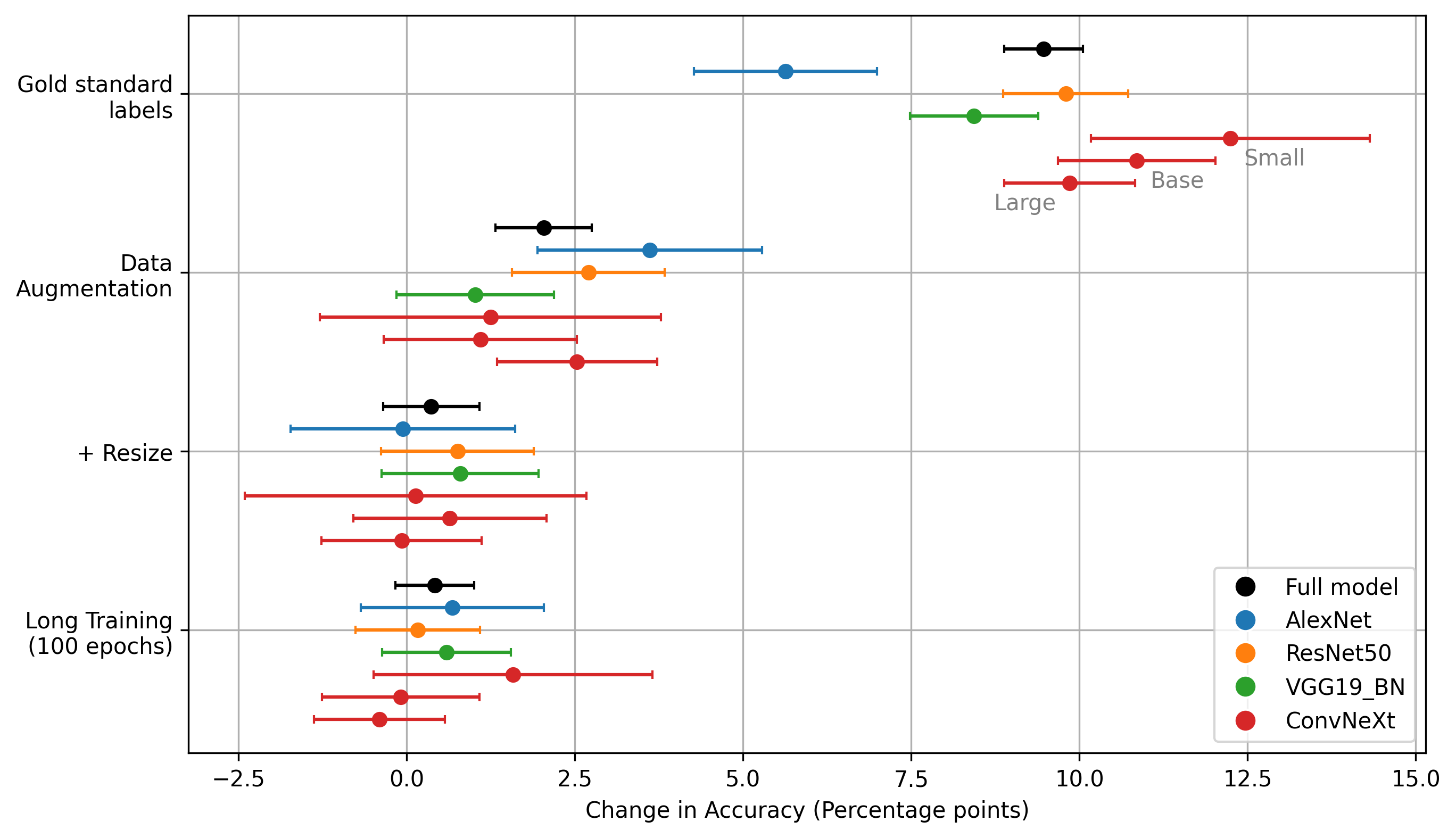} 

}

\caption{Effect of hyperparameter variation on accuracy across experiments. Dots represent point estimates with error bars showing 95 percent confidence intervals. Reference categories are: interviewer-scored data, no augmentation, and 50 epochs of training.}\label{fig:hyper-imp}
\end{figure}

Although this is a rather arbitrary comparison, it is clear that the
ConvNeXt models perform well and substantially better than human
interviewers, even on a surprisingly small dataset. In future research
we will use much larger datasets, which may require a re-evaluation, but
for the time being we will use ConvNeXt as our model architecture of
choice for further analysis.

\hypertarget{impact-of-hyperparameters}{%
\subsection{Impact of hyperparameters}\label{impact-of-hyperparameters}}

To analyse the impact of the hyperparameters, we estimated linear
regression models for all CNN model architectures individually, with
prediction accuracy as the dependent variable and the hyperparameters as
the independent variables. In addition, we estimated a model on the full
dataset of experimental results where we also controlled for the CNN
model architecture to get a sense of the architecture-independent effect
of the hyperparameters.

Figure \ref{fig:hyper-imp} shows that doubling the training time from 50
to 100 epochs had no significant effect on training accuracy, which may
be due to the models quickly adapting to such a small dataset size. Data
augmentation had a relatively modest but still significant effect of 2
percentage points in the full model. This effect varied considerably
between model architectures, with AlexNet benefiting the most. The
addition of the resize variant did not significantly improve prediction
accuracy. As seen in the previous figure, the effect of the gold
standard is significant at around 9.5 percentage points. AlexNet only
gets a comparatively small boost of 5.5 percentage points, while
ConvNeXt improves by about 12 percentage points compared to when trained
on the interviewer-scored labels. However, both have a relatively wide
confidence interval.

\hypertarget{classification-of-errors}{%
\subsection{Classification of errors}\label{classification-of-errors}}

As mentioned above, not all cubes are equally difficult to score. In
particular, the partially correct cubes are problematic as they are
often incorrectly scored by interviewers (see Figure \ref{fig:gsintc}).
We therefore looked more closely at the errors made by the models
trained on the gold standard data.

\begin{figure}

{\centering \includegraphics[width=0.5\linewidth]{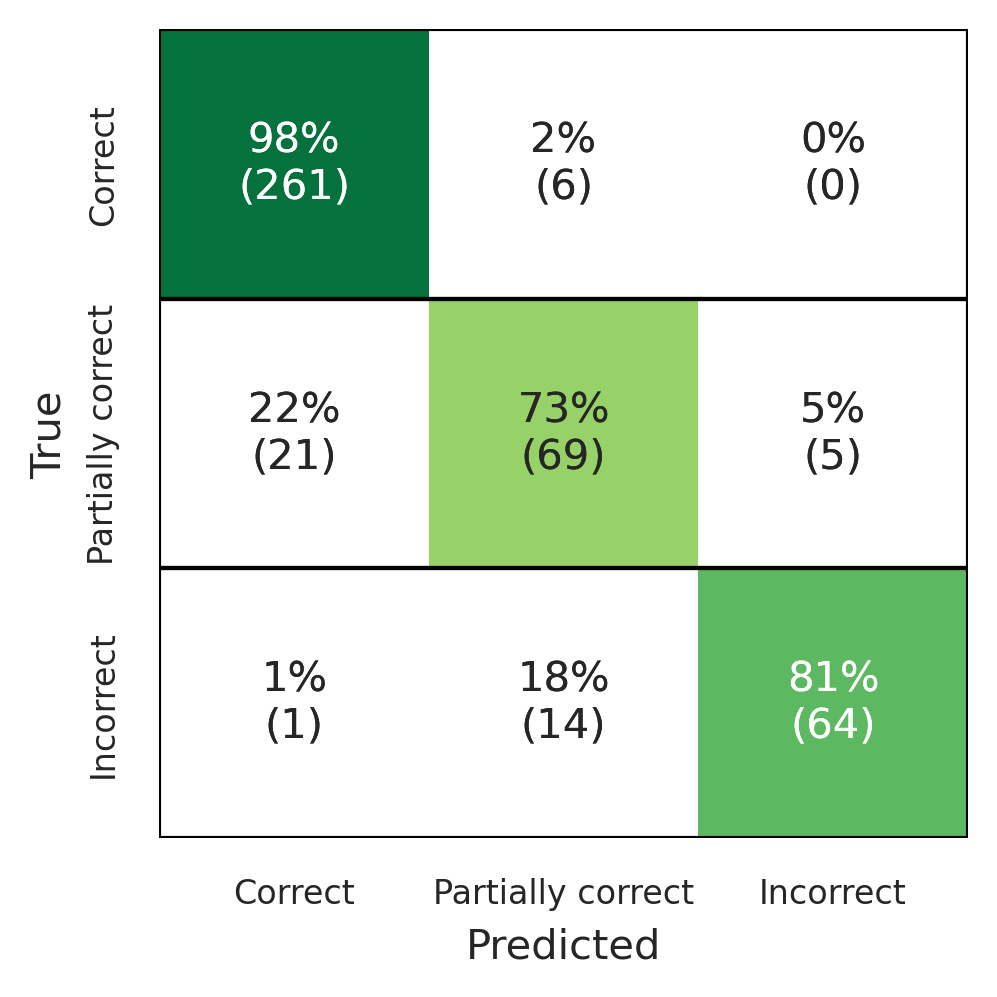} 

}

\caption{Confusion matrix of the best ConvNeXt Base model trained with data augmentation for 100 epochs on the gold standard data. Percentages in rows show the proportion of model predictions in each class with correct predictions highlighted. Parantheses contain absolute numbers.}\label{fig:gs-mod}
\end{figure}

Figure \ref{fig:gs-mod} shows the confusion matrices for the ConvNeXt
Base model. The model performs best on the correct cubes, with an
accuracy of 98\%. Incorrectly drawn cubes are classified as such with an
accuracy of 81\%. As we saw with the interviewers, the partially correct
cubes are the most difficult to classify, with an accuracy of 73\%. This
is certainly better than the interviewer performance, but still much
worse than the other classes.

The model we used for comparison outperforms the interviewers in all
classes by a considerable margin, but still struggles with the partially
correct cubes. One reason for this may be that, as an intermediate
category, it is inherently ambiguous and thus harder to classify than
the more clear-cut correct and incorrect categories. This also makes it
difficult to assign correct labels during the gold standard scoring
process. The fact that the scoring rules are so vague exacerbates the
problem, which is why it seems crucial to rely on expert judgement in
the future. The small sample size of partially correct cubes in the gold
standard data may also be a factor. This is probably also true for the
incorrect class, where accuracy is higher but not outstanding.

\hypertarget{automated-scoring-in-production}{%
\subsection{Automated scoring in
production}\label{automated-scoring-in-production}}

\begin{figure}

{\centering \includegraphics[width=1\linewidth]{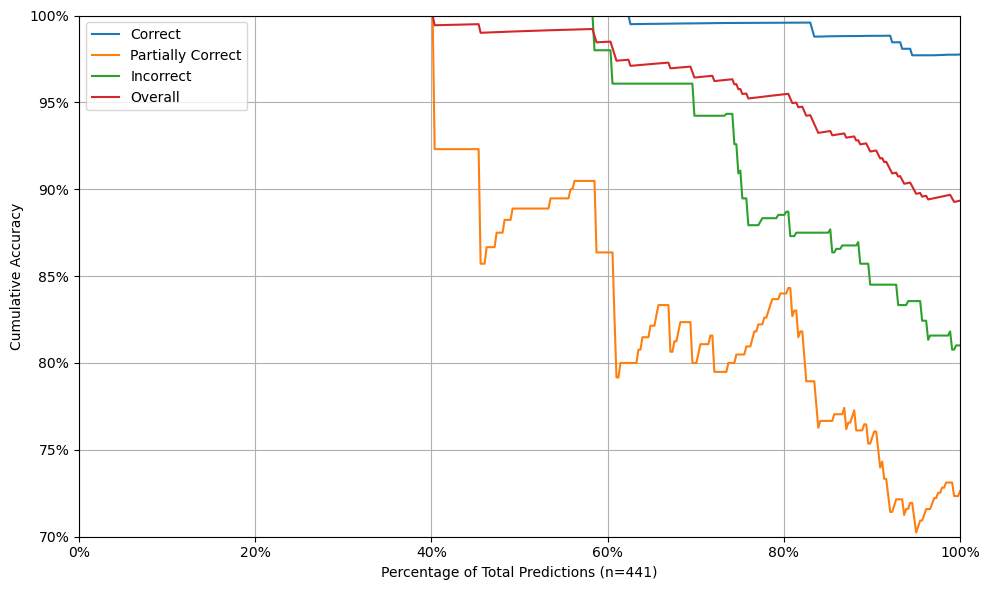} 

}

\caption{Production plot of cumulative accuracy over prediction confidence per class. Predictions are based on the best ConvNeXt Base model, trained with data augmentation for 100 epochs on the gold standard data.}\label{fig:prod-plot}
\end{figure}

If we are going to use automated scoring based on model predictions to
improve the quality of the data in the SHARE data releases made
available to substantive researchers, we need to know what proportion of
the cubes can be automatically labelled while still maintaining a level
of accuracy that we consider acceptable. To get an idea of this
relationship, we ordered the cubes by the confidence of the prediction,
i.e.~the probability of the most likely class, on the x-axis, with the
most confident on the left and the least confident on the right. We then
plotted this against the cumulative accuracy of the model predictions on
the y-axis.\footnote{see also Bethmann et al. (2014), where we used a
  similar approach for automated coding of occupations in survey data}.
To distinguish between different types of error, we also plotted the
cumulative accuracy of the correct, partially correct and incorrect
cubes separately. The result is shown in Figure \ref{fig:prod-plot}.

What we see in the plot is that for predictions with very high
confidence, the accuracy is also very high, but that the accuracy drops
off rapidly as the confidence decreases along the x-axis. On the far
right we end up with the same percentages we saw in Figure
\ref{fig:gs-mod}, which makes sense as this would be the case if we just
let the model automatically score all the cubes. If we want higher
accuracy we need to move to the left, but this comes at the cost of
having to score more cubes manually. For example, if we wanted to have
an overall accuracy of about 95\%, we would need to manually score about
20\% of the cubes and let the model score 80\%, which is still a
significant reduction in the scoring burden. If we are willing to accept
an overall accuracy of 90\%, we can reduce the scoring burden to about
5\%. Depending on how we value the errors in the different classes, we
can place more emphasis on the class-specific curves. For example, if we
wanted to have an accuracy of about 85\% for the partially correct
cubes, we would need to score about 40\% of the cubes manually and let
the model score 60\%.

\begin{figure}

{\centering \includegraphics[width=1\linewidth]{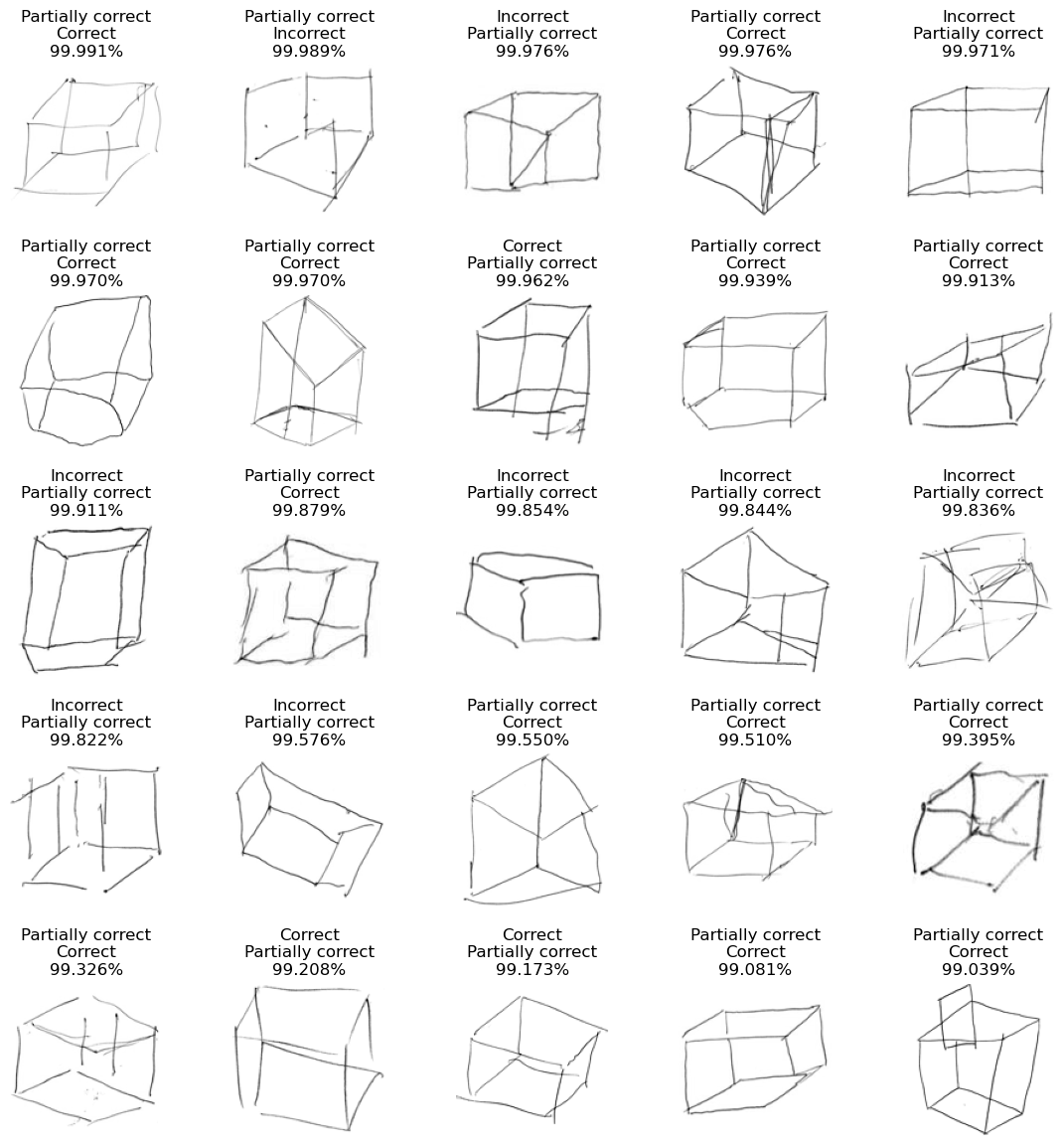} 

}

\caption{Misclassified cubes with highest confidence. True class, predicted class and prediction confidence are given above each drawing.}\label{fig:misclass}
\end{figure}

There are some caveats to these graphs. They only give an estimate of
the expected accuracy, as confidence and accuracy are based on the
model's prediction, and as we can see, the model itself is far from
perfect. This is indicated not only by the accuracies being (well) below
100\%, but also by the jaggedness of the downward slope, which could be
due to errors in labelling the data, or the model struggling to classify
certain types of cubes. In either case, the model made an incorrect
classification even though it was very confident about the prediction
(see Figure \ref{fig:misclass} for a few examples). This is obviously a
problem if the model is to be used in production, and should be
mitigated by improving both the training data and the model (see some
suggestions in the next section).

\hypertarget{discussion}{%
\section{Discussion}\label{discussion}}

In conclusion, we have shown that it is indeed possible to use deep
learning to automate the scoring of drawing tests in cognitive
assessments collected in surveys. In this proof of concept, we trained
several different CNN models on two types of datasets: one labelled with
scores provided by survey interviewers during fieldwork, and another
labelled with an in-house `gold standard' scoring procedure. The models
in the ConvNeXt family averaged over 85\% overall prediction accuracy in
our experiments when trained on the gold standard data, meaning they
outperformed human interviewers by around 5 percentage points. This is a
promising result, especially considering that it can be achieved without
any further human intervention, thus reducing the burden on the
interviewer while improving the quality of the scoring at virtually no
additional cost.

\hypertarget{limitations}{%
\subsection{Limitations}\label{limitations}}

We would be reluctant to use the current models in production at this
stage. The main reason for this is that the models, like the
interviewers, are still struggling to score the partially correct cubes.
They have not yet achieved an acceptable level of accuracy in this
class. This is probably due to the small number of cases in the training
dataset (about 350) and possibly also to ambiguities in the gold
standard scores. At the same time, we were quite surprised to achieve
this level of performance with a training dataset of only about 1,300
cases in total after subtracting the validation dataset.

\hypertarget{increasing-the-quantity-of-training-data}{%
\subsection{Increasing the quantity of training
data}\label{increasing-the-quantity-of-training-data}}

In terms of the volume of cube drawings, we have already collected
around 55,000 additional recording booklets from SHARE waves eight and
nine in several countries. Once these have been scanned and processed,
we will have a much larger dataset on which to train our models. The
caveat here is that so far we only have interviewer scores for these
cubes, which, as we have shown, are not a particularly good basis for
training. We therefore need to score these cubes using our gold standard
procedure.

As this is a very time-consuming process, we will need to adapt the
procedure to make it more efficient. To do this, we will look at
automatic pre-labelling of the cubes using the models trained on the
current dataset, as well as active learning approaches such as
uncertainty sampling (e.g. Cohn et al., 1994; Settles, 2012). This will
allow us to focus on the more difficult, and therefore useful, cases and
thus speed up the process.

While the larger dataset is likely to mitigate some of the detrimental
effects of class imbalance on model performance, in particular the small
number of partially correct cubes, we will also want to explore the
possibility of using data augmentation to generate more of these cases.
A particularly interesting approach could be to use diffusion models to
generate synthetic data, which have recently been shown to be very
effective in improving performance in image classification tasks (e.g.
Azizi et al., 2023; Sehwag, 2023).

\hypertarget{improving-the-quality-of-the-scoring}{%
\subsection{Improving the quality of the
scoring}\label{improving-the-quality-of-the-scoring}}

We will also need to rethink the assumption that our `gold standard'
scores are not only more consistent, and therefore easier to train on,
but actually more accurate than interviewer scores. This means that for
at least a representative subset of the gold standard scores we will
need to obtain expert judgements from, for example, clinical
psychologists, something that we might actually want to incorporate into
our use of active learning approaches.

It will also be interesting to look more closely at the scoring rules
and the scoring process itself. As we have seen, the partially correct
cubes are particularly problematic and the scoring rules are not very
clear. It seems questionable whether the scoring rules in their current
form are actually a good way of communicating the scoring criteria to
the scorers. At the moment there seems to be a lot of hidden knowledge
that scorers have to acquire through experience, which may work well if
the scorers are true experts, such as clinical psychologists, who can
apply their knowledge gained through clinical training to the scoring
process. This does not seem feasible for ordinary survey interviewers or
lay office scorers, who would have to rely much more on scoring rules,
training examples and expert supervision. We will therefore look more
closely at what different types of scorers actually do, what their
inter-rater reliabilities are, and how this relates to quality criteria
such as expert ratings, other test scores and diagnoses. One way to
inform these analyses is to use the trained CNN models to visualise the
parts of the drawings that are most important for classification with
techniques such as Grad-CAM (Selvaraju et al., 2016). This will give us
a better idea of how the scoring rules currently work and how they could
be improved, at least if we want to continue using lay scorers to score
cognitive drawing tests in the near future.

Respondents draw cubes in many different ways and the deviations from
the template pictures are very varied. These deviations are often not
random, but follow certain patterns that appear in many drawings, for
reasons that may include situational factors, personal style, practice,
cultural differences, physical impairments or, of course, different
levels of cognitive functioning in the visuospatial domain. At the same
time, it is often not clear, especially to a lay scorer, whether a
particular deviation should affect the score or not. The scoring rules
are far too short and vague to capture this level of detail. It can
therefore be useful to use dimensionality reduction or clustering
techniques to get an idea of the types of deviation patterns present in
the distribution (Ester et al., 1996; Maaten \& Hinton, 2008; McInnes et
al., 2018). The combination of expert judgement and model-based accuracy
predictions could then be used to identify relevant clusters, for
example because they are ambiguously scored or because they are a
particularly salient indicator of cognitive impairment. This could help
to guide the labelling process beyond the creation of more complex but
generic rules (as is done in other cognitive assessments) and expert
supervision of scoring.

\hypertarget{improving-the-models}{%
\subsection{Improving the models}\label{improving-the-models}}

There is also plenty of room for improvement on the modelling side,
including but not limited to more hyperparameter tuning.

The ImageNet pre-training weights we used, as provided by the
fastai/timm default option, may not be the best, or even the most
obvious, basis for our fine-tuning. The dataset consists mainly of
photographs, which are probably not ideal for pre-training the layers of
a CNN tasked with classifying human-drawn sketches of very simple
objects. We will therefore explore more suitable pre-training datasets
in future research, such as ImageNet-Sketch (Wang et al., 2019), TU
Berlin Sketch Dataset (Eitz et al., 2012), QuickDraw (Ha \& Eck, 2017),
or perhaps just the cumulated set of cube, infinity loop, and clock
drawings from our own data.

In terms of model architectures, we will also explore the use of Vision
Transformers (Dosovitskiy et al., 2020), which are currently considered
state-of-the-art for many image classification tasks. We expect them to
be a viable option, especially as dataset size increases, potentially
challenging the performance of the best CNNs.

\hypertarget{future-work}{%
\subsection{Future work}\label{future-work}}

As a future step, we will score the other two drawings from the ACE-III
test, the overlapping infinity loops and the clock drawing, using the
best practices learned from scoring the cube drawings. We will also
explore the possibility of using the models trained on the cube drawings
to pre-label the other drawings, which should speed up the process
considerably.

An interesting avenue to explore is transfer learning from the cube
drawings to other drawings. This could be done by simply fine-tuning the
cube-trained models with the training datasets of, say, the infinity
loops or the clocks. The rationale would be that the features learned
during training on the cube dataset are more easily tuned to the
classification task of other similar drawings than the original
pre-training weights based on, for example, ImageNet. Similarly, we
could explore transfer learning from our models to cognitive tests in
other studies, such as those used in HCAP, which include very similar
drawing tests. This could be particularly interesting since, for
example, SHARE-HCAP is a sub-sample of about 2,500 respondents from the
SHARE sample, and fine-tuning a good model for a similar task might make
it possible to achieve production-level prediction accuracy even with
such small training datasets.

Finally, we also look forward to working on making our data and models
available to the scientific community. We plan to make the drawings and
the corresponding scores available as open datasets. We also want to
release the models and the code used to train them, so that other
researchers can replicate our results and build on them in their own
research. Researchers will also be able to use the models to score their
own drawings, which we hope will lead to more standardised and
consistent scoring of cognitive drawing tests in future research.

\hypertarget{acknowledgments}{%
\section*{Acknowledgments}\label{acknowledgments}}
\addcontentsline{toc}{section}{Acknowledgments}

This paper uses data from SHARE Wave 8 (DOI: 10.6103/SHARE.w8.800) see
Börsch-Supan et al. (2013) for methodological details. The SHARE data
collection has been funded by the European Commission, DG RTD through
FP5 (QLK6-CT-2001-00360), FP6 (SHARE-I3: RII-CT-2006-062193, COMPARE:
CIT5-CT-2005-028857, SHARELIFE: CIT4-CT-2006-028812), FP7 (SHARE-PREP:
GA N°211909, SHARE-LEAP: GA N°227822, SHARE M4: GA N°261982, DASISH: GA
N°283646) and Horizon 2020 (SHARE-DEV3: GA N°676536, SHARE-COHESION: GA
N°870628, SERISS: GA N°654221, SSHOC: GA N°823782, SHARE-COVID19: GA
N°101015924) and by DG Employment, Social Affairs \& Inclusion through
VS 2015/0195, VS 2016/0135, VS 2018/0285, VS 2019/0332, and VS
2020/0313. Additional funding from the German Ministry of Education and
Research, the Max Planck Society for the Advancement of Science, the
U.S. National Institute on Aging (U01\_AG09740-13S2, P01\_AG005842,
P01\_AG08291, P30\_AG12815, R21\_AG025169, Y1-AG-4553-01, IAG\_BSR06-11,
OGHA\_04-064, HHSN271201300071C, RAG052527A) and from various national
funding sources is gratefully acknowledged (see www.share-project.org).

\hypertarget{references}{%
\section*{References}\label{references}}
\addcontentsline{toc}{section}{References}

\hypertarget{refs}{}
\begin{CSLReferences}{1}{0}
\leavevmode\vadjust pre{\hypertarget{ref-Amini2021}{}}%
Amini, S., Zhang, L., Hao, B., Gupta, A., Song, M., Karjadi, C., Lin,
H., Kolachalama, V. B., Au, R., \& Paschalidis, I. C. (2021). An
artificial intelligence-assisted method for dementia detection using
images from~the clock drawing test. \emph{Journal of Alzheimer's Disease
: JAD}, \emph{83}(7, 2), 581--589.
\url{https://doi.org/10.1017/S1355617720000144}

\leavevmode\vadjust pre{\hypertarget{ref-Azizi2023}{}}%
Azizi, S., Kornblith, S., Saharia, C., Norouzi, M., \& Fleet, D. J.
(2023). \emph{Synthetic data from diffusion models improves ImageNet
classification}. \url{https://doi.org/10.48550/ARXIV.2304.08466}

\leavevmode\vadjust pre{\hypertarget{ref-Bethmann2014}{}}%
Bethmann, A., Schierholz, M., Wenzig, K., \& Zielonka, M. (2014).
Automatic coding of occupations. \emph{Proceedings of Statistics Canada
Symposium 2014. Beyond Traditional Survey Taking: Adapting to a Changing
World}.
\url{https://www.statcan.gc.ca/sites/default/files/media/14291-eng.pdf}

\leavevmode\vadjust pre{\hypertarget{ref-Binaco2020}{}}%
Binaco, R., Calzaretto, N., Epifano, J., McGuire, S., Umer, M., Emrani,
S., Wasserman, V., Libon, D. J., \& Polikar, R. (2020). Machine learning
analysis of digital clock drawing test performance for differential
classification of mild cognitive impairment subtypes versus alzheimer's
disease. \emph{Journal of the International Neuropsychological Society},
\emph{26}(7), 690--700.
https://doi.org/\url{https://doi.org/10.1017/S1355617720000144}

\leavevmode\vadjust pre{\hypertarget{ref-BoerschSupan2022}{}}%
Börsch-Supan, A. (2022). \emph{Survey of health, ageing and retirement
in europe (SHARE) wave 8}. SHARE-ERIC.
\url{https://doi.org/10.6103/SHARE.W8.800}

\leavevmode\vadjust pre{\hypertarget{ref-BoerschSupan2013}{}}%
Börsch-Supan, A., Brandt, M., Hunkler, C., Kneip, T., Korbmacher, J.,
Malter, F., Schaan, B., Stuck, S., \& Zuber, S. (2013). Data resource
profile: The survey of health, ageing and retirement in europe (SHARE).
\emph{International Journal of Epidemiology}, \emph{42}(4), 992--1001.
\url{https://doi.org/10.1093/ije/dyt088}

\leavevmode\vadjust pre{\hypertarget{ref-BoerschSupan2005}{}}%
Börsch-Supan, A., \& Jürges, H. (Eds.). (2005). \emph{The survey of
health, ageing and retirement in europe -- methodology}. MEA.
\url{https://share-eric.eu/fileadmin/user_upload/Methodology_Volumes/Methodology_2005.pdf}

\leavevmode\vadjust pre{\hypertarget{ref-Cohn1994}{}}%
Cohn, D., Atlas, L., \& Ladner, R. (1994). Improving generalization with
active learning. \emph{Machine Learning}, \emph{15}(2), 201--221.
\url{https://doi.org/10.1007/bf00993277}

\leavevmode\vadjust pre{\hypertarget{ref-Deng2009}{}}%
Deng, J., Dong, W., Socher, R., Li, L.-J., Li, K., \& Fei-Fei, L. (2009,
June). ImageNet: A large-scale hierarchical image database. \emph{2009
IEEE Conference on Computer Vision and Pattern Recognition}.
\url{https://doi.org/10.1109/cvpr.2009.5206848}

\leavevmode\vadjust pre{\hypertarget{ref-Dosovitskiy2020}{}}%
Dosovitskiy, A., Beyer, L., Kolesnikov, A., Weissenborn, D., Zhai, X.,
Unterthiner, T., Dehghani, M., Minderer, M., Heigold, G., Gelly, S.,
Uszkoreit, J., \& Houlsby, N. (2020). \emph{An image is worth 16x16
words: Transformers for image recognition at scale}.
\url{https://doi.org/10.48550/ARXIV.2010.11929}

\leavevmode\vadjust pre{\hypertarget{ref-eitz2012hdhso}{}}%
Eitz, M., Hays, J., \& Alexa, M. (2012). How do humans sketch objects?
\emph{ACM Trans. Graph. (Proc. SIGGRAPH)}, \emph{31}(4), 44:1--44:10.
\url{https://cybertron.cg.tu-berlin.de/eitz/pdf/2012_siggraph_classifysketch.pdf}

\leavevmode\vadjust pre{\hypertarget{ref-ester1996density}{}}%
Ester, M., Kriegel, H.-P., Sander, J., \& Xu, X. (1996). A density-based
algorithm for discovering clusters in large spatial databases with
noise. \emph{Kdd}, \emph{96}(34), 226--231.

\leavevmode\vadjust pre{\hypertarget{ref-Folstein1975}{}}%
Folstein, M. F., Folstein, S. E., \& McHugh, P. R. (1975). "Mini-mental
state". \emph{Journal of Psychiatric Research}, \emph{12}(3), 189--198.
\url{https://doi.org/10.1016/0022-3956(75)90026-6}

\leavevmode\vadjust pre{\hypertarget{ref-Ha2017}{}}%
Ha, D., \& Eck, D. (2017). \emph{A neural representation of sketch
drawings}. \url{https://doi.org/10.48550/ARXIV.1704.03477}

\leavevmode\vadjust pre{\hypertarget{ref-He2015}{}}%
He, K., Zhang, X., Ren, S., \& Sun, J. (2015). \emph{Deep residual
learning for image recognition}.
\url{https://doi.org/10.48550/ARXIV.1512.03385}

\leavevmode\vadjust pre{\hypertarget{ref-info11020108}{}}%
Howard, J., \& Gugger, S. (2020). Fastai: A layered API for deep
learning. \emph{Information}, \emph{11}(2).
\url{https://doi.org/10.3390/info11020108}

\leavevmode\vadjust pre{\hypertarget{ref-Hsieh2013}{}}%
Hsieh, S., Schubert, S., Hoon, C., Miosh, E., \& Hodges, J. (2013).
Validation of the addenbrooke's cognitive examination III in
frontotemporal dementia and alzheimer's disease. \emph{Dementia and
Geriatric Cognitive Disorders}, \emph{36}(3--4), 242--250.
\url{https://doi.org/10.1159/000351671}

\leavevmode\vadjust pre{\hypertarget{ref-Hu2022}{}}%
Hu, M., Murphey, Y., Wang, S., Qin, T., Zhao, Z., Gonzalez, R.,
Freedman, V., \& Zahodne, L. (2022). Exploring the use of machine
learning to improve dementia detection: Automating coding of the
clock-drawing test. \emph{Innovation in Aging}, \emph{6}, 382--382.
\url{https://doi.org/10.1093/geroni/igac059.1508}

\leavevmode\vadjust pre{\hypertarget{ref-Krizhevsky2012}{}}%
Krizhevsky, A., Sutskever, I., \& Hinton, G. E. (2012). ImageNet
classification with deep convolutional neural networks. In F. Pereira,
C. J. Burges, L. Bottou, \& K. Q. Weinberger (Eds.), \emph{Advances in
neural information processing systems} (Vol. 25). Curran Associates,
Inc.
\url{https://proceedings.neurips.cc/paper_files/paper/2012/file/c399862d3b9d6b76c8436e924a68c45b-Paper.pdf}

\leavevmode\vadjust pre{\hypertarget{ref-Langa2019}{}}%
Langa, K. M., Ryan, L. H., McCammon, R. J., Jones, R. N., Manly, J. J.,
Levine, D. A., Sonnega, A., Farron, M., \& Weir, D. R. (2019). The
health and retirement study harmonized cognitive assessment protocol
project: Study design and methods. \emph{Neuroepidemiology},
\emph{54}(1), 64--74. \url{https://doi.org/10.1159/000503004}

\leavevmode\vadjust pre{\hypertarget{ref-Liu2022}{}}%
Liu, Z., Mao, H., Wu, C.-Y., Feichtenhofer, C., Darrell, T., \& Xie, S.
(2022). \emph{A ConvNet for the 2020s}.
\url{https://doi.org/10.48550/ARXIV.2201.03545}

\leavevmode\vadjust pre{\hypertarget{ref-Maaten2008}{}}%
Maaten, L. van der, \& Hinton, G. (2008). Visualizing data using t-SNE.
\emph{Journal of Machine Learning Research}, \emph{9}(86), 2579--2605.
\url{http://jmlr.org/papers/v9/vandermaaten08a.html}

\leavevmode\vadjust pre{\hypertarget{ref-Malhotra2015}{}}%
Malhotra, R., Haaland, B. A., Chei, C.-L., Chan, A., Malhotra, C., \&
Matchar, D. B. (2015). Presence of and correction for interviewer error
on an instrument assessing cognitive function of older adults.
\emph{Geriatrics \& Gerontology International}, \emph{15}(3), 372--380.
\url{https://doi.org/10.1111/ggi.12331}

\leavevmode\vadjust pre{\hypertarget{ref-McInnes2018}{}}%
McInnes, L., Healy, J., \& Melville, J. (2018). \emph{UMAP: Uniform
manifold approximation and projection for dimension reduction}.
\url{https://doi.org/10.48550/ARXIV.1802.03426}

\leavevmode\vadjust pre{\hypertarget{ref-Nasreddine2005}{}}%
Nasreddine, Z. S., Phillips, N. A., Bédirian, V., Charbonneau, S.,
Whitehead, V., Collin, I., Cummings, J. L., \& Chertkow, H. (2005). The
montreal cognitive assessment, MoCA: A brief screening tool for mild
cognitive impairment. \emph{Journal of the American Geriatrics Society},
\emph{53}(4), 695--699.
\url{https://doi.org/10.1111/j.1532-5415.2005.53221.x}

\leavevmode\vadjust pre{\hypertarget{ref-Plank2022}{}}%
Plank, B. (2022). \emph{The 'problem' of human label variation: On
ground truth in data, modeling and evaluation}.
\url{https://doi.org/10.48550/ARXIV.2211.02570}

\leavevmode\vadjust pre{\hypertarget{ref-Say2022}{}}%
Say, M. J., \& O'Driscoll, C. (2022). Inter-rater variability in scoring
of addenbrooke's cognitive examination-third edition (ACE-III)
protocols. \emph{Applied Neuropsychology: Adult}, 1--5.
\url{https://doi.org/10.1080/23279095.2022.2083964}

\leavevmode\vadjust pre{\hypertarget{ref-sehwag_minimal_diffusion_2023}{}}%
Sehwag, V. (2023). Minimal-diffusion: A minimal yet resourceful
implementation of diffusion models (along with pretrained models +
synthetic images for nine datasets). \emph{GitHub Repository}.
\url{https://github.com/VSehwag/minimal-diffusion}

\leavevmode\vadjust pre{\hypertarget{ref-Selvaraju2016}{}}%
Selvaraju, R. R., Cogswell, M., Das, A., Vedantam, R., Parikh, D., \&
Batra, D. (2016). Grad-CAM: Visual explanations from deep networks via
gradient-based localization. \emph{International Journal of Computer
Vision}, \emph{128}(2), 336--359.
\url{https://doi.org/10.1007/s11263-019-01228-7}

\leavevmode\vadjust pre{\hypertarget{ref-Settles2012}{}}%
Settles, B. (2012). Uncertainty sampling. In \emph{Synthesis lectures on
artificial intelligence and machine learning} (pp. 11--20). Springer
International Publishing.
\url{https://doi.org/10.1007/978-3-031-01560-1_2}

\leavevmode\vadjust pre{\hypertarget{ref-Simonyan2014}{}}%
Simonyan, K., \& Zisserman, A. (2014). \emph{Very deep convolutional
networks for large-scale image recognition}.
\url{https://doi.org/10.48550/ARXIV.1409.1556}

\leavevmode\vadjust pre{\hypertarget{ref-Wagner2021}{}}%
Wagner, M., \& Douhou, S. (2021). Cognitive measures. In M. Bergmann \&
A. Börsch-Supan (Eds.), \emph{SHARE wave 8 methodology: Collecting
cross-national survey data in times of COVID-19} (pp. 43--46). MEA, Max
Planck Institute for Social Law; Social Policy.
\url{https://share-eric.eu/fileadmin/user_upload/Methodology_Volumes/SHARE_Methodenband_WEB_Wave8_MFRB.pdf}

\leavevmode\vadjust pre{\hypertarget{ref-Wang2019}{}}%
Wang, H., Ge, S., Xing, E. P., \& Lipton, Z. C. (2019). Learning robust
global representations by penalizing local predictive power.
\emph{NeurIPS 2019}. \url{https://doi.org/10.48550/ARXIV.1905.13549}

\leavevmode\vadjust pre{\hypertarget{ref-West2016}{}}%
West, B. T., \& Blom, A. G. (2016). Explaining interviewer effects: A
research synthesis. \emph{Journal of Survey Statistics and Methodology},
\emph{5}(2), 175--211. \url{https://doi.org/10.1093/jssam/smw024}

\leavevmode\vadjust pre{\hypertarget{ref-rw2019timm}{}}%
Wightman, R. (2019). PyTorch image models. In \emph{GitHub repository}.
\url{https://github.com/rwightman/pytorch-image-models}; GitHub.
\url{https://doi.org/10.5281/zenodo.4414861}

\leavevmode\vadjust pre{\hypertarget{ref-WHO2017}{}}%
World Health Organization. (2017). \emph{Global action plan on the
public health 2005 - 2007}.
\url{https://www.who.int/publications/i/item/9789241513487}

\leavevmode\vadjust pre{\hypertarget{ref-WHO2021}{}}%
World Health Organization. (2021). \emph{Global status report on the
public health response to dementia}.
\url{https://www.who.int/publications/i/item/9789240033245}

\leavevmode\vadjust pre{\hypertarget{ref-Youn2021}{}}%
Youn, Y. C., Pyun, J.-M., Ryu, N., Baek, M. J., Jang, J.-W., Park, Y.
H., Ahn, S.-W., Shin, H.-W., Park, K.-Y., \& Kim, S. Y. (2021). Use of
the clock drawing test and the rey-osterrieth complex figure test-copy
with convolutional neural networks to predict cognitive impairment.
\emph{Alzheimer's Research \& Therapy}, \emph{13}(1), 85.
\url{https://doi.org/10.1186/s13195-021-00821-8}

\end{CSLReferences}

\end{document}